\documentclass[twocolumn,showpacs,preprintnumbers,amsmath,amssymb]{revtex4-1}

\usepackage{graphicx}% Include figure files
\usepackage{dcolumn}% Align table columns on decimal point
\usepackage{bm}% bold math
\usepackage{epstopdf}
\usepackage{amssymb}
\usepackage[caption=false]{subfig} 
\begin{document}

%\preprint{APS/123-QED}

\title{Room evacuation through two contiguous
exits} %  

\author{I.M.~Sticco}
 \affiliation{Departamento de F\'\i sica, Facultad de Ciencias 
Exactas y Naturales, Universidad de Buenos Aires,\\
 Pabell\'on I, Ciudad Universitaria, 1428 Buenos Aires, Argentina.}
 %Lines break automatically or can be forced with \\
\author{G.A.~Frank}
 \affiliation{Universidad Tecnol\'ogica Nacional, Facultad Regional Buenos 
Aires, Av. Medrano 951, 
 1179 Buenos Aires, Argentina.}
 %Lines break automatically or can be forced with \\
\author{C.O.~Dorso}%
 \email{codorso@df.uba.ar}
\affiliation{Departamento de F\'\i sica, Facultad de Ciencias 
Exactas y Naturales, Universidad de Buenos Aires,}%
 \affiliation{Instituto de F\'\i sica de Buenos Aires,
 Pabell\'on I, Ciudad Universitaria, 1428 Buenos Aires, Argentina.}%
\date{\today}% It is always \today, today,
             %  but any date may be explicitly specified

\begin{abstract}
Current regulations demand that at least two exits should be available 
for a safe evacuation during a panic situation. Although the ``faster is 
slower'' effect is expected to take place near the exits, the evacuation 
time will improve because of the additional exits. However, rooms having 
contiguous doors not always reduce the leaving time as expected. We 
investigated the relation between the doors separation and the 
evacuation performance. We found that there exists a separation distance 
range that does not really improve the evacuation time, or it can even worsen 
the process performance. To our knowledge, no attention has been given to this 
issue in the literature. This work reports how the pedestrians dynamics differ 
when the separation distance between two exit doors changes and how this 
affects the overall performance.
\end{abstract}


\maketitle

\section{\label{introduction}Introduction}

The practice of providing two doors for emergency evacuation can be traced back to the last Qing dynasty in China (1644-1911 AD). A mandatory regulation established that large buildings had to provide two fire exits \cite{cheng}. This kind of regulations upgraded to current standard codes with detailed specifications on the exits position, widths and separations \cite{OSHA,FLO}.  
\medskip

Current regulations claim that the minimum door width should be 0.813~m while 
the maximum door-leaf should not exceed 1.219~m \cite{FLO,FLO2}. If more than 
two doors are required, the distance between two of then must be at least 
one-half or one-third of the room diagonal distance. But, no special 
requirements apply to the rest of the doors, regardless the fact that they 
should not be simultaneously blocked \cite{FLO,FLO2}.  
\medskip

The rulings leave some space for placing the extra openings (\emph{i.e.} those 
above two exits) at an arbitrary separation distance. Thus, it is possible to 
place a couple of doors on the same side of the room. The special case of two 
contiguous doors has been examined throughout the literature 
\cite{kirchner1,perez1,daoliang1,huanhuan1}. 
\medskip

Kirchner and Schadschneider studied the pedestrians evacuation process through 
two contiguous door using a cellular automaton model \cite{kirchner1}. The 
agents were able to leave the room under increasing panic situations for 
behavioural patterns varying from individualistic pedestrians to strongly 
coupled pedestrians moving like a \emph{herd}. The evacuation time was found 
to be independent of the separation distance for the individualistic 
pedestrians in a panic situation. But if the pedestrians were allowed to move 
like a herd, an increasing evacuation time for small separation lengths (less 
than 10 individuals size) was reported. 
\medskip

The total number of pedestrians leaving the room per 
unit time actually showed a slow-down for separation distances smaller than 
four door widths \cite{perez1}. This slow-down was identified as a disruptive 
interference effect due to pedestrians crossing in each other's path.  The 
threshold of four door widths ($4\,d_w$) corresponds to the distance separation 
necessary to distinguish two independent groups of pedestrians, each one
surrounding the nearest door. 
\medskip

Researchers called the attention on the fact that no matter how separated 
the two contiguous doors are placed, the overall performance does not improve 
twice with respect to a single exit (of the same width). This 
effect is attributed to some sort of pedestrians interference  \cite{perez1}. 
\medskip 

Although the above results were obtained for very narrow doors (\emph{i.e.} 
single individual width), further investigation showed that they also apply 
to doors allowing two simultaneous leaving pedestrians. However, this does 
not hold for a room with a single door \cite{daoliang1}. In this case, it is 
true that the 
mean flux of evacuating people increases with an increasing door width, but the 
flux per unit width decreases \cite{daoliang1}.    
\medskip

The separation distance of the two doors may worsen the evacuation performance 
if the doors are placed close to the end of the wall, that is, near each 
corner of the room. People can get in contact with other 
walls, loosing evacuation efficiency \cite{kirchner1,daoliang1}.  
\medskip

More detailed investigation on cellular automata evacuation processes showed 
that the evacuation performance depends on five distances: the total width 
of the openings (that is, adding the widths of each door), the doors 
separation distance, the width difference between the two doors, and the 
distance to the nearest corner \cite{huanhuan1}.    
\medskip

The total width of the openings can improve the evacuation time for any 
separation distance between doors, if both doors have the same width. 
The door separation, however, controls the optimal location for 
these exits. A rough rule states that the doors separation distance 
$d_g$ should equal $L-4\,d_w$, where $L$ is the room length 
\cite{huanhuan1}.  
\medskip

The optimal location rule agrees with the slow-down phenomenon for 
small values of $d_g$. It also agrees with the worsening of the evacuation 
performance for doors close to a corner. But this kind of worsening may also 
appear for other reasons, since it has been suggested that the relatively longer 
traveling distance of the pedestrians to the doors also plays an important 
role.   
\medskip

The two doors configuration does not need to be symmetric along the wall. 
Asymmetry causes delays depending on the width difference between the doors and 
their relative position. It has been shown that placing the wider door in the 
middle of the wall, and another one at the corner (with reduced width), causes 
an improvement in the evacuation process \cite{huanhuan1}. 
\medskip

Our investigation focuses only symmetric configurations with equal sized doors. 
As opposed to the above mentioned literature, we examine the evacuation 
dynamics by means of the Social Force Model (SFM). An overview of this model 
can be found in Section \ref{background}.  
\medskip

In Sections \ref{background} and \ref{simulations} the single door 
configuration is revisited. Its purpose is to make easier the understanding of 
the two-doors configuration for very small separation distances $d_g$.  
\medskip

In Sections \ref{simulations} and \ref{results} we examine the effects of 
increasing the $d_g$ until the clogging areas close to each door become almost 
independent.  
\medskip

Section \ref{conclusions} resumes the pedestrians behavioural patterns, and its 
consequences on the evacuation performance, for the different door 
separation scenarios.

\section{\label{background}Background}

\subsection{The Social Force Model}

The ``social force model'' (SFM) deals with the pedestrians behavioural 
pattern in a crowded environment. The basic model states that the 
pedestrians motion is controlled by three kind of forces: the ``desire force'', 
the ``social force'' and the ``granular force''. The three are very different in 
 nature, but enter into an equation of motion as follows
\medskip

\begin{equation}
m_i\,\displaystyle\frac{d\mathbf{v}^{(i)}}{dt}(t)=\mathbf{f}_d^{(i)}
(t)+\displaystyle\sum_{j}\displaystyle\mathbf{f}_s^{(ij)}(t)+\displaystyle\sum_{
j}\mathbf{f}_g^{(ij)}(t)\label{eqn_1}
\end{equation}

\noindent where $m_i$ is the mass of the pedestrian $i$, and $\mathbf{v}_i$ is 
its corresponding velocity. The subscript $j$ represents all other pedestrians 
(excluding $i$) and the walls. $\mathbf{f}_d$, $\mathbf{f}_s$ and 
$\mathbf{f}_g$ are the desire force, the social force and the granular force, 
respectively.
\medskip

The desire force resembles the pedestrian's own desire to go to a specific 
place \cite{Helbing1}. He (she) needs to accelerate (decelerate) from his (her) 
current velocity, in order to achieve its own willings. As he (she) reaches the 
velocity that makes him (her) feel comfortable, no further acceleration 
(deceleration) is required. This velocity is the ``desired velocity'' of the 
pedestrian $\mathbf{v}_d(t)$. The expression for $\mathbf{f}_d$ in 
Eq.~(\ref{eqn_2}) handles this issue.  
\medskip

\begin{equation}
\left\{\begin{array}{lcl}
        \mathbf{f}_d^ {(i)}(t) & = & m_i\,\displaystyle\frac{\mathbf{v}_d^
{(i)}(t)-\mathbf{v}_i(t)}{\tau} \\
        & & \\
\mathbf{f}_s^{(ij)} & = & A_i\,e^{(r_{ij}-d_{ij})/B_i}\mathbf{n}_{ij}\\
        & & \\
\mathbf{f}_g^{(ij)} &= &\kappa\,(r_{ij}-d_{ij})\,\Theta(r_{ij}-d_{ij})\,\Delta
\mathbf{v}_{ij}\cdot\mathbf{t}_{ij} \\
       \end{array}\right.\label{eqn_2}
\end{equation}

$\tau$ means a relaxation time. Further details on each parameter con be found 
in Refs.~\cite{Helbing1,Dorso1,Dorso2,Dorso3,Dorso4}.
\medskip

Notice that the desired velocity $\mathbf{v}_d$ has magnitude $v_d$ and points 
to the desired place in the direction $\hat{\mathbf{e}}_d$. Thus, $v_d$ 
represents his (her) state of anxiety, white $\hat{\mathbf{e}}_d$ indicates the 
place where he (she) is willing to go. We assume, for simplicity, that 
$v_d$ remains constant during an evacuation process, but $\hat{\mathbf{e}}_d$ 
changes according to the current position.   
\medskip

The social force $\mathbf{f}_s$ corresponds to the tendency of each individual 
to keep some space between him and other pedestrian, or, between him and the 
walls \cite{Helbing4}. The $\mathbf{f}_s$ expressed in Eq.~(\ref{eqn_2}) 
depends on the inter-pedestrian distance $d_{ij}$. The magnitude 
$r_{ij}=r_i+r_j$ is the sum of the pedestrian's radius, while $A_i$ and $B_i$ 
are two fixed parameters ($r_j=0$ for the interaction with the wall). Thus, 
$\mathbf{f}_s$ is a repulsive monotonic force that resembles the pedestrian 
feelings for preserving his (her) \textit{private sphere} 
\cite{Helbing1,Helbing4}. 
\medskip

The granular force $\mathbf{f}_g$ appearing in Eq.~(\ref{eqn_1}) represents the 
sliding friction between contacting people (or between people  and walls). Its 
expression can be seen in Eq.~(\ref{eqn_2}). It is assumed to be a linear 
function of the relative (tangential) velocities $\Delta
\mathbf{v}_{ij}\cdot\mathbf{t}_{ij}$ of the contacting individuals. The 
function $\Theta(r_{ij}-d_{ij})$ returns the argument value if $r_{ij}>d_{ij}$, 
while $\kappa$ is a fixed parameter (see 
Refs.~\cite{Helbing1,Dorso1,Dorso2,Dorso3,Dorso4}).
\medskip

\subsection{\label{human}Clustering structures}

The time delays during an evacuation process are related to clustering people 
as explained in Refs.~\cite{Dorso1,Dorso2}. Groups of contacting pedestrians can 
be defined as the set of individuals that for any member of the group (say, 
$i$) there exists at least another member belonging to the same group ($j$) for 
whom $d_{ij}<r_i+r_j$. This kind of structure is called a \textit{human 
cluster}. 
\medskip

From all human clusters appearing during the evacuation process, those that 
are simultaneously in contact with the walls on both sides of the exit are 
the ones that possibly \textit{block} the way out. Thus, we are interested 
in the minimum number of contacting pedestrians belonging to this 
\textit{blocking cluster} that are able to link both sides of the exit. We call 
this minimalistic group as a \textit{blocking structure}. Any blocking 
structure is supposed to work as a barrier for the pedestrians in behind.
\medskip

\subsection{\label{pressure}The local pressure on the pedestrians}

Recall that the social force model (SFM) deals with the pedestrians desire and 
their private space preservation. Although the desire force 
$\mathbf{f}_d$ and the social force $\mathbf{f}_s$ are not exactly 
``physical'' quantities (\textit{i.e.} not pair-wise), the movement 
equation remains valid. Therefore, it can be derived from the virial relation 
that \cite{lion}

\begin{equation}
 \bigg\langle\displaystyle\sum_{i=1}^N\displaystyle\frac{p_i^2}{m_i} + 
\displaystyle\sum_{i=1}^N 
\mathbf{r}_i\cdot\mathbf{f}_i\bigg\rangle=-3\mathcal{PV}\label{eqn_3}
\end{equation}

\noindent for the set of $N$ pedestrians inside a volume $\mathcal{V}$. $p_i$  
and $\mathbf{f}_i$ are the momentum and total force acting on the individual $i$ 
(excluding the interaction with the walls). $\langle\cdot\rangle$ corresponds to 
the mean value along time. The right hand side $-3\mathcal{PV}$ defines the 
global pressure on the surface enclosing the volume $\mathcal{V}$.  
\medskip

The local pressure on a single pedestrian (say, $i$) corresponds to the forces 
(per unit area) acting on him due to the surrounding pedestrians. Following 
Ref.~\cite{lion} we can define the ``social pressure 
function'' $P_i$ as
\medskip

\begin{equation}
3P_iV_i=\displaystyle\frac{p_i^2}{m_i} + \frac{1}{2}
\displaystyle\sum_{j=1}^{N-1}
\mathbf{r}_{ij}\cdot\mathbf{f}_s^{(ij)}\label{eqn_4}
\end{equation}

\noindent where $V_i$ is the volume enclosing the pedestrian $i$ and 
$\mathbf{r}_{ij}=\mathbf{r}_{i}-\mathbf{r}_j$. Notice that the inner product 
$\mathbf{r}_{ij}\cdot\mathbf{f}_s^{(ij)}$ is always positive for repulsive 
feelings and equals the scalar product $d_{ij}f_s^{(ij)}$.  
\medskip 

The second term in Eq.~(\ref{eqn_3}) can be split into the sum of the inner
products $\mathbf{r}_i\cdot\mathbf{f}_d$ (desire), 
$\mathbf{r}_i\cdot\mathbf{f}_s$ (social) and $\mathbf{r}_i\cdot\mathbf{f}_g$ 
(granular). Actually, the sum of the social product depends on the 
inter-pedestrian distance $d_{ij}f_s^{(ij)}$, while the granular one does not 
play a role because of orthogonality  
($\mathbf{r}_{ij}\cdot\mathbf{f}_g^{(ij)}=0$). Consequently, the virial 
relation (\ref{eqn_3}) reads 
\medskip

\begin{equation}
 \displaystyle\sum_{i=1}^N\langle3P_iV_i 
\rangle=-3\mathcal{PV}-\displaystyle\sum_{i=1}^N \langle
\mathbf{r}_i\cdot\mathbf{f}_d^{(i)}\rangle\label{eqn_5}
\end{equation}

We should remark that Eq.~(\ref{eqn_5}) holds either if the pedestrians are in 
contact or not. The ``social pressure function'' $P_i$ makes possible 
for the individuals to change their behavioural pattern when they come too 
close to each other or to the walls. 
\medskip 

In Appendix~\ref{bulk_pressure} we apply Eq.~(\ref{eqn_5}) to a simple 
example and compare it to our results. 

\section{\label{simulations}Numerical simulations}

\subsection{\label{numerical_geometry} Geometry and process simulation}

\begin{figure}
\includegraphics[width=\columnwidth]{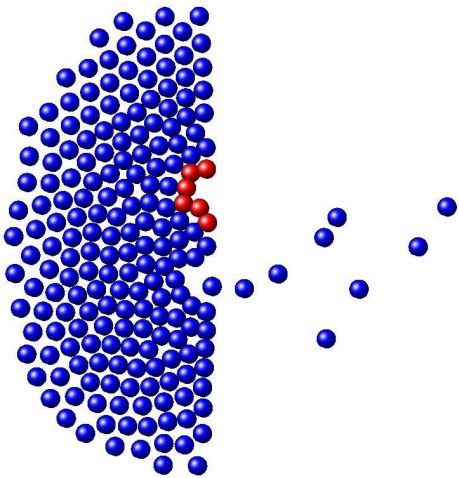}
\caption{\label{fig:19} Snapshot of an evacuation process from a 
$20\,\mathrm{m}\times20\,$m room, with two doors. In red we can see a blocking 
structure around the upper door. The desired velosity was $v_d=4\,$m/s.  }
%  done by sticco
\end{figure}

We simulated different evacuation processes for room sizes of 20~m $\times$ 
20~m, 30~m $\times$ 30~m and 40~m $\times$ 40~m. The rooms had one or two exit 
doors on the same wall, as shown in Fig.~\ref{fig:19}. The doors were placed 
symmetrically from the mid position of the wall, in order to avoid corner 
effects. Both doors had also the same width. 
\medskip

At the beginning of the process, the pedestrians were all equally separated 
in a square arrangement. The occupancy density was set to 0.6~people/m$^2$,  
close to the allowed limiting values in current regulations \cite{mysen}. 
They all had random velocities resembling a Gaussian distribution with null 
mean value. The pedestrians were willing to go to the nearest exit. Thus, all 
the pedestrians had the desired velocity $\mathbf{v}_d$ pointing to the same 
exit door if only one door was available, or to the nearest door if two exits 
were available.  
\medskip

In order to focus on the effects due to dual exits, we only allowed the 
pedestrians to move individualistically, that is, neither leaderships nor 
herding behaviors were present during the evacuation process. At any time, the 
pedestrians knew the doors location and tried to escape by their own.  \medskip

The simulations were supported by {\sc Lammps} molecular 
dynamics simultator with parallel computing capabilities \cite{plimpton}. The 
time integration algorithm followed the velocity Verlet scheme with a time step 
of $10^{-4}\,$s. All the necessary parameters were set to the same values as in 
previous works (see Refs.~\cite{Dorso3,Dorso4}). It was assumed that all the 
individuals had the same radius ($r_i=0.3$~m) and weight ($m_i=70$~kg). We ran 
30 processes for each panic situation, in order to get enough data for mean 
values computation. 
\medskip

Although the {\sc Lammps} simulator has the most common built-in functions, 
neither the social force $\mathbf{f}_s$  nor the desire force $\mathbf{f}_d$ 
were available. We implemented special modules (with parallel computing 
compatibilities) for the $\mathbf{f}_s$ and $\mathbf{f}_d$ computations. 
These computations were checked over with previous computations. 
\medskip

The pedestrian's desired direction $\hat{\mathbf{e}}_d$ was updated at each 
time step. After leaving the room, they continued moving away. No re-entering 
mechanism was allowed. 
\medskip

\subsection{\label{numerical_data}Measurements conditions}

Simulations were run in the same way as in Refs. \cite{Dorso3,Dorso4}.  Each 
process started with all the individuals inside the room. The measurement period 
lasted until 80\% of the occupants left the room. If this condition could not be 
fulfilled within the first 3000~s, the process was stopped. Data was 
recorded at time intervals of $0.05\tau$ (cf. Eq.~(\ref{eqn_2}a).
\medskip

The simulations ran from relaxed situations ($v_d<2\,$m/s) to very stressing 
rushes ($v_d=6\,$m/s). We registered the individuals positions and 
velocities for each evacuation process. Thus, we were able to compute the 
``social pressure'' through out the process and to trace the pedestrians 
behavioural pattern.

\section{\label{results}Results}

\subsection{\label{faster_is_slower}The faster is slower effect}

\begin{figure}
\includegraphics[width=\columnwidth]{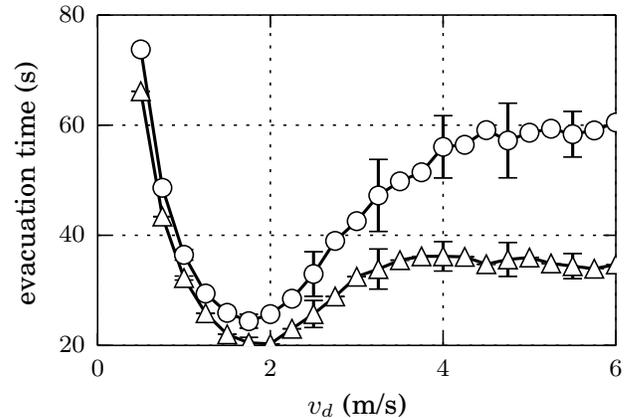}
\caption{\label{fig:7} Mean evacuation time for 160 individuals (seconds) 
vs. the pedestrian's desired velocity (m/s). Two doors were available for 
leaving the room (see text for details). Mean values were computed from 30 
evacuation processes. Each door was $d_w=1.2$~m width. The desired velocity 
was $v_d=4\,$m/s. Two situations are shown:  $\bigtriangleup$ corresponds to 
the null separation distance between doors, meaning a single door of $2L$ 
width. $\bigcirc$ corresponds to the 1~m separation distance between doors.   }
% done with fig7_version0.py 
\end{figure}

As a starting point, we checked over the ``faster is slower'' effect for the 
the room with two doors on the same wall. Fig.~\ref{fig:7} shows the 
recorded evacuation time when the doors are separated a distance of $d_g=1\,$m 
and when no separation exists at all ($d_g=0$). The latter means a single 
opening with width equal to two doors. Both cases (with or without separation) 
exhibit a change in their corresponding derivatives. Thus, the ``faster is 
slower'' effect is achieved following the same qualitative response as the one 
found in previous works for rooms with a single exit \cite{Helbing1,Dorso1}.
\medskip

The evacuation time for separated doors in Fig.~\ref{fig:7} is always above the 
time required to evacuate the pedestrians through the single opening 
(\emph{i.e.} null separation). For $v_d=6\,$m/s, the single opening improves 
the evacuation performance in half of the time that demands the $d_g=1\,$m 
separation configuration. Other separation distances (not shown) exhibit the 
same qualitative pattern as the example presented in Fig.~\ref{fig:7}. 
Therefore, it is clear that while the total width of the opening 
remains unchanged, splitting this width into to symmetric exits may affect 
significantly the evacuation performance. 
\medskip

The results shown in Fig.~\ref{fig:7} are in agreement with the 
conclusions reported in Ref.~\cite{perez1}. Any door separation distance less 
than  $4d_w$ (see caption in Fig.~\ref{fig:7}) produce a ``slow-down'' 
in the evacuation performance. However, we did not come to consensus on the 
reasons of this slow-down, as expressed in Ref.~\cite{perez1}. After running 
some animations, we realized that the ``disruptive interference effect'' 
mentioned there could not be the main reason for the increase in the 
evacuation delays. 
\medskip

We made further inquiries on the $d_g=0$ and $d_g>0$ scenarios. The 
former is investigated in Section \ref{single_double}, while the latter is left 
to Section \ref{door_seperation}. 
\medskip

\subsection{\label{single_double}The single door vs. the null separation}

Recall that the $d_g=0$ scenario corresponds to one door with an opening equal 
to twice its width (see Section~\ref{faster_is_slower}). According to 
Ref.~\cite{huanhuan1}, this widening is expected to affect the evacuation 
performance. 
\medskip

\subsubsection{\label{null_gap_data}The stop-and-go process}

\begin{figure*}[!htbp]
\subfloat[Opening of $d_w=1.2$~m width (single door exit).\label{fig:9}]{
\includegraphics[width=\columnwidth]{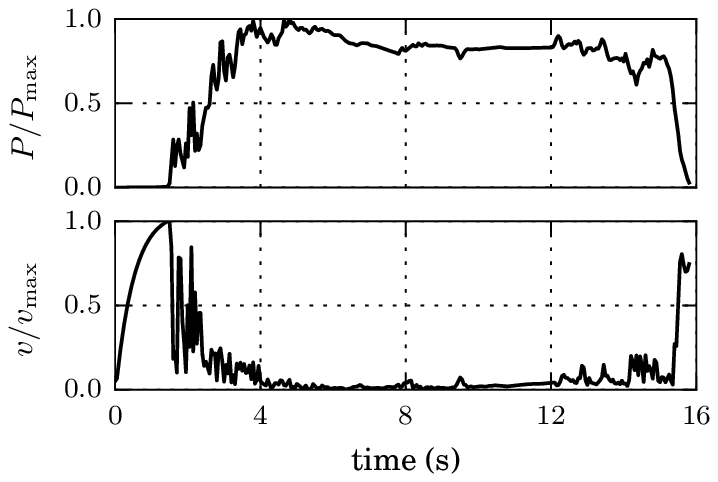}
% done with fig4_version0.py 
}\hfill
\subfloat[Opening of $3d_w=3.6$~m width (null separation). \label{fig:8}]{
\includegraphics[width=\columnwidth]{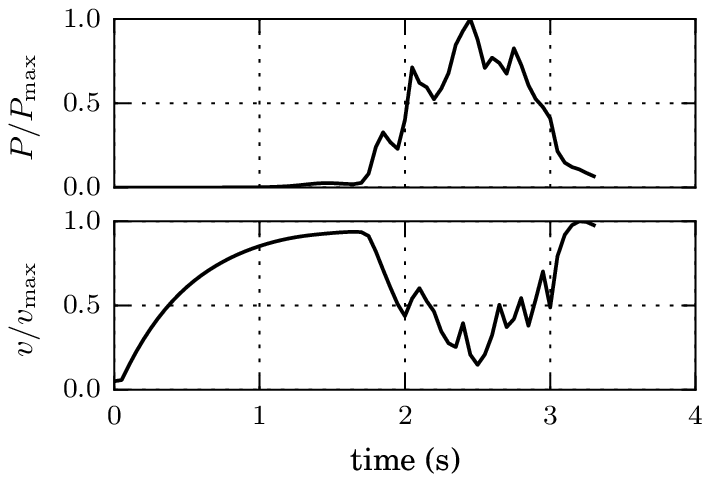}
% done with fig8_version0.py 
}
\caption{\label{fig:8and9} Normalized pressure and velocity on a single 
pedestrian during an evacuation process. Data was recorded from the 
the initial position at $x=12.35$~m and $y=8.45$~m, until the individual left 
the room ($x>20$~m).  The pedestrians desired velocity was $v_d=4\,$m/s. Two 
situations are shown: (a) evacuation through a single door of width 
$d_w=1.2$~m. 
(b) evacuation through an opening of $3d_w=3.6$~m (resembling the null 
separation situation for two doors of width $1.5\,L$).} 
\end{figure*}

Fig.~\ref{fig:8and9} illustrates on how the evacuation performance can be 
improved as the opening becomes wider.  Fig.~\ref{fig:9} corresponds to 
the single door, while  Fig.~\ref{fig:8} corresponds to the null 
separation situation (wider opening). We can see the (normalized) pressure and 
velocity evolution of a tagged pedestrian that rushes from 
$(x,y)=(12.35\,\mathrm{m},8.45\,\mathrm{m})$ 
to the exit. His (her) evacuation time is reduced by five when the opening 
widens by three (see Fig.~\ref{fig:8and9}). This is in partial agreement with 
Ref.~\cite{daoliang1} since the widening improves the flux of evacuating people, 
but we cannot assure that the flux per unit width decreases, as the authors in 
Ref.~\cite{daoliang1} suggest.  
\medskip

The tagged pedestrian in Fig.~\ref{fig:8and9} increases his (her) velocity 
towards an asymptotic value at the beginning of the processes. This value 
corresponds to the desired velocity $v_d=4\,$m/s. But close to $t=2\,$s, the 
pedestrian suddenly stops because of the clogging around the exit. Clogging is 
also responsible for the pressure increase, as shown in both 
Fig.~\ref{fig:9} and Fig.~\ref{fig:8} (for details on how the pressure was 
computed, see bellow in this section). Notice, however, that any further 
fluctuation of the pressure acting on the tagged pedestrian corresponds to an 
inverse fluctuation on the velocity. Thus, the pedestrian is able to reach the 
exit following a stop-and-go process. 
\medskip

The instantaneous pressure acting on a single pedestrian can be computed from 
Eq.~\ref{eqn_4}. If we neglect the momentum $p_i$ for any slow moving 
pedestrian, then $3P_iV_i$ in Eq.~\ref{eqn_4} equals the sum of the products 
$d_{ij}f_s^{(ij)}$. Actually, not all the inter-pedestrian distances 
$d_{ij}$ need to be computed since only the first neighbors contribution to 
the social force become relevant. The magnitude $3P_iV_i$ is the one 
represented in Fig.~\ref{fig:9} and Fig.~\ref{fig:8} (normalized by 
$3\mathrm{P}_\mathrm{max}V_i$).
\medskip

The maximum pressure values $3\mathrm{P}_\mathrm{max}V_i$ in 
Fig.~\ref{fig:9} and Fig.~\ref{fig:8} are 8783~N.m and 7103 N.m, respectively. 
The corresponding mean pressure values (after the first $2\,$s) are 80\% and 
55\% of the respective maximum values. This means that although the reported 
maximum pressure is quite similar in both situations, there is a noticeable 
bias in the mean values. The wider opening seems to release pressure from time 
to time. Consequently, the stop-and-go processes are somehow different for the 
single door with respect to the $d_g=0$ situation (the wider opening). 
\medskip

\subsubsection{\label{null_gap_patterns}The pressure and stream patterns}

\begin{figure*}[!htbp]
\subfloat[Mean pressure contour lines. The scale bar on the right is 
expressed in N.m units (see text for details). Level colors can be seen in the 
on-line version only.  
\label{fig:3}]{\includegraphics[width=\columnwidth]{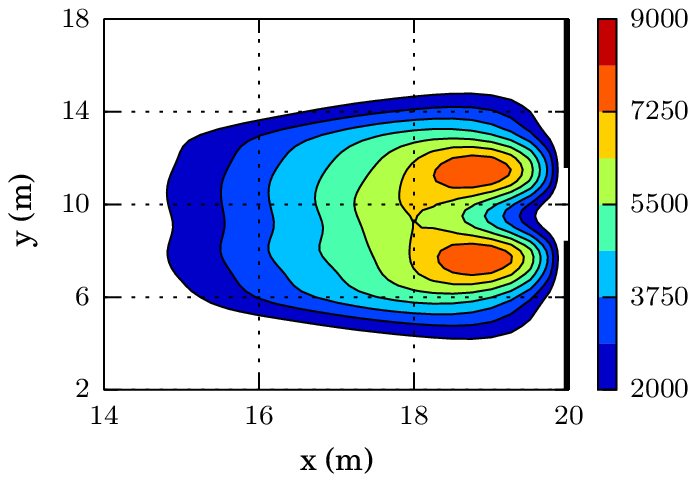}
% done with fig3_version0.py
}\hfill
\subfloat[Mean stream lines. The lines connect the normalized
velocity field ($v/v_\mathrm{max}$). The arrows indicate the stream 
direction.\label{fig:5}]{
\includegraphics[width=\columnwidth]{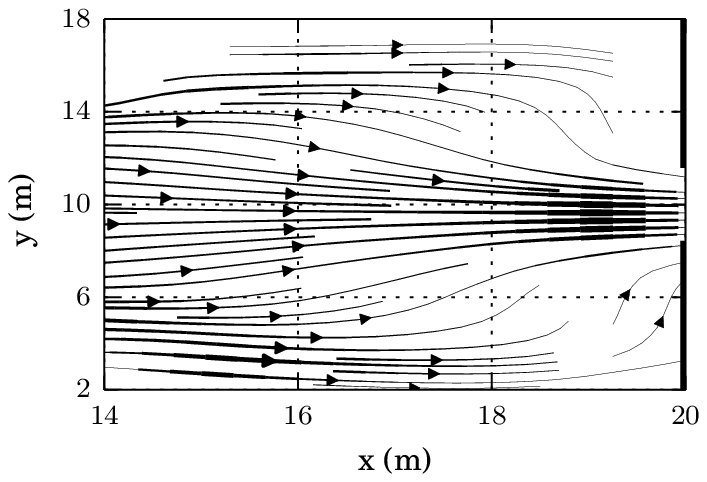}
% done with fig5_version0.py
}
\caption{\label{fig:3and5}Mean pressure and stream lines computed from 30 
evacuation processes until 100 pedestrians left the room 
($20\,\mathrm{m}\times20\,\mathrm{m}$ size). Data was recorded on a square grid 
of $1\,\mathrm{m}\times1\,\mathrm{m}$ and then splined to get smooth curves. 
The thick black lines at $x=20$~m represent the walls on the right of the room. 
There is only one opening of $3d_w=3.6$~m width (null separation distance 
between 
doors of width $3L/2$). The pedestrian's desired velocity was $v_d=4\,$m/s.  }
\end{figure*}

For a better understanding on how the pedestrians are (intermittently) released 
from high pressures in the wide opening situation, we pictured the whole scene 
into a pressure contour map and a mean stream path map for all the individuals. 
Fig.~\ref{fig:3} shows the pressure levels ($3P_iV_i$) for the clogging area. 
The warm colors are associated to high pressure values. These values 
are close to the corresponding maximum pressure values (not shown). Thus, the 
warm regions define the places where the pedestrians slow down most of the 
time. They are expected to get released only for short periods of time. On the 
contrary, the regions represented in cold colors (low mean pressure) are those 
where the individuals are supposed to get released for longer time periods. 
\medskip

Fig.~\ref{fig:5} represents the mean stream lines during the evacuation 
process. It completes the stop-and-go picture since it exhibits the 
released paths for leaving the room. Notice that the stream lines pass 
through the low pressure regions. That is, it can be seen in Fig.~\ref{fig:5} 
that the stream lines gather along the middle of the clogging area, 
where ``cold'' pressure colors can be found (cf. Fig.~\ref{fig:3}). The 
``warm'' pressure colors are placed on the sides of this region.    
\medskip

We checked over the trajectory of the tagged pedestrian represented in 
Fig.~\ref{fig:8} and we observed that he (she) managed to get out of the room 
through the path where the stream lines gather. Thus, Fig.~\ref{fig:8} 
resembles the stop-and-go process for the pedestrians passing through the 
middle of the clogging area, that is, along the low pressure (middle)
region. The pedestrians on the sides of this region (high pressure region) are 
expected to slow down since Fig.~\ref{fig:5} shows no stream lines to the exit. 
\medskip

\begin{figure*}[!htbp]
\subfloat[Opening of $d_w=1.2$~m width (single door exit).\label{fig:4}]{
\includegraphics[width=\columnwidth]{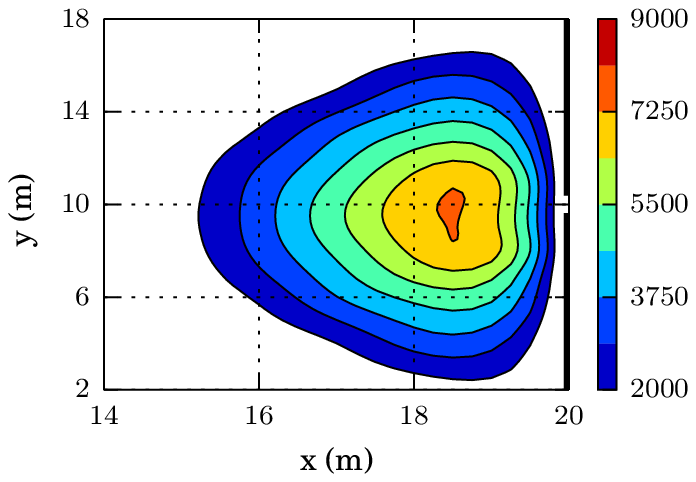}
% done with fig4_version0.py 
}\hfill
\subfloat[Opening of $2d_w=2.4$~m width (null separation distance between doors 
of 
width $L$). \label{fig:2}]{
\includegraphics[width=\columnwidth]{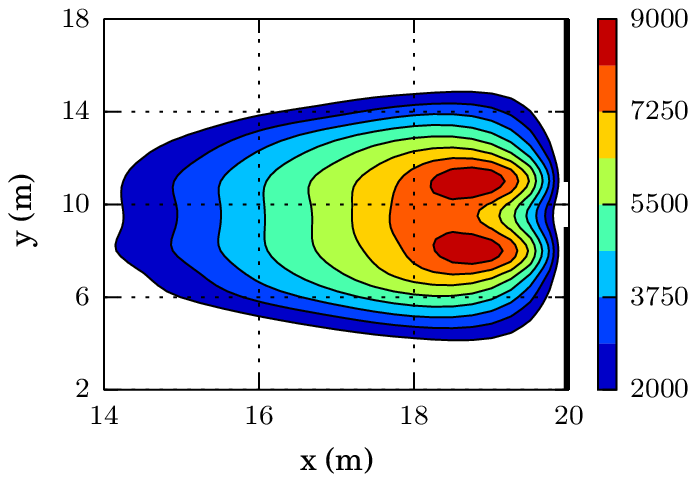}
% done with fig2_version0.py 
}
\caption{\label{fig:2and4} Mean pressure contour lines computed from 30 
evacuation processes until 100 pedestrians left the room 
($20\,\mathrm{m}\times20\,\mathrm{m}$ size). The scale bar on the right is 
expressed in N.m units (see text for details). The thick black lines at 
$x=20$~m represent the walls on the right of the room. The pedestrian's desired 
velocity was $v_d=4\,$m/s. The contour lines were computed on a square grid of 
$1\,\mathrm{m}\times1\,\mathrm{m}$ and then splined to get smooth curves. Level 
colors can be seen in the on-line version only.} 
\end{figure*}

Recalling the results in Fig.~\ref{fig:9} for the same tagged individual as in 
Fig.~\ref{fig:8}, we realize that the single door scene is likely 
to differ from the $d_g=0$ situation since both patterns (for the same 
individual) do. Thus, we examined the pressure contour map for the single door 
and for an opening of twice the single door width. The results are shown in 
Fig.~\ref{fig:2and4}. Fig.~\ref{fig:2} exhibits a similar pressure map pattern 
as Fig.~\ref{fig:3}, but the single door pressure map in Fig.~\ref{fig:4} does 
not. For the single door situation, we do not observe the lower pressure 
pathway in the middle of the clogging area. Instead, high pressure 
is acting on the pedestrians, as shown in the (normalized) pressure evolution 
in Fig.~\ref{fig:9}. The corresponding velocity evolution (Fig.~\ref{fig:9}) 
informs that the pedestrians in this region experience a slow down. 
\medskip

At this stage of the investigation we are able to point out a few conclusions. 
The widening of the single door increases the pedestrian's flux, as asserted in 
Ref.~\cite{daoliang1}. In the single door situation, the pedestrians experience 
a slow down close to the exit. These time delays have been associated to 
blocking structures (see Refs.~\cite{Dorso1,Dorso2}) and causes the 
pressure acting on the nearby individuals to rise. Fig.~\ref{fig:4} resembles 
this situation. However, as the opening widens (\emph{i.e.} the null separation 
situation), the pressure pattern changes qualitatively (see Fig.~\ref{fig:3} 
and Fig.~\ref{fig:2}), allowing the pedestrians in the middle of the clogging 
area to make a pathway to the exit. This pathway corresponds to the breaking of 
the blocking structures.   
\medskip

\subsection{\label{door_seperation} Separated doors}

The second relevant distance mentioned in Ref.~\cite{huanhuan1} is the doors 
separation distance. We will fix the door width $d_w=1.2\,$m and focus only on 
the evacuation processes from a room with two doors symmetrically placed on 
the same side of the room. 
\medskip

\begin{figure}
\includegraphics[width=\columnwidth]{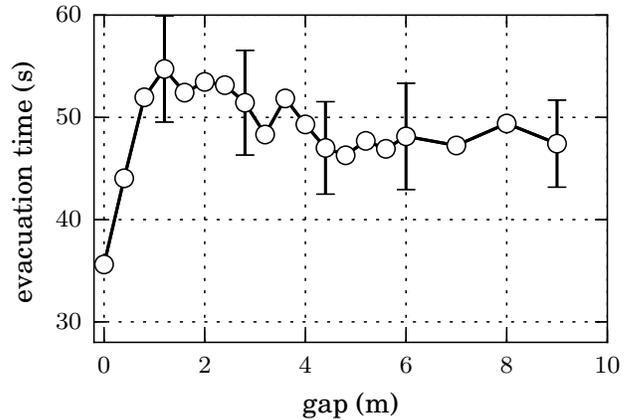}
\caption{\label{fig:13} Mean evacuation time for 225 pedestrians (room of 
$20\times20$~m size) as a function of the doors separation distance. Mean 
values were computed from 30 evacuation processes until 160 pedestrians left 
the room. Each door was $d_w=1.2$~m width for non-vanishing gaps. The null gap 
means a single door of $2L$ width. The desired velocity was $v_d=4\,$m/s. }
% done with fig13_version0.py 
\end{figure}

It has been shown in Fig.~\ref{fig:7} that separating the doors a distance 
$d_g=1\,$m worsens the evacuation performance. We  further explored this 
worsening by increasing $d_g$ at steps of $0.5\,$m, starting from the null 
separation distance. Fig.~\ref{fig:13} shows the mean evacuation time and the 
corresponding error bars (indicating the $\pm\sigma$ limits). The desired 
velocity was set to $v_d=4\,$m/s, achieving the ``faster is slower'' scenario. 
\medskip

The evacuation time as a function of $d_g$ shown in Fig.~\ref{fig:13} is one 
of our main results. The worsening in the evacuation performance rises to a 
maximum value, and surprisingly, its derivative changes sign for $d_g>1\,$m. 
Thus, $d_g=1\,$m appears to be the worst evacuation scenario for 
the $20\,\mathrm{m}\times20\,\mathrm{m}$ room with 225 individuals and two 
doors of $d_w=1.2\,$m each (see Fig.~\ref{fig:13}).  
\medskip

Notice that Fig.~\ref{fig:13} is not in complete agreement with the literature 
\cite{perez1,huanhuan1}. As outlined in Section~\ref{introduction}, it has been 
argued that the optimal separation distance (gap) $d_g$ should be bounded 
between $4d_w$ and $L-4\,d_w$ ($L$ means the room side, while $d_w$ is the doors 
width). An inspection of Fig.~\ref{fig:13} confirms this, although very small 
values of $d_g$ can also enhance the evacuation time. Thus, it is not 
completely true that the ``disruptive interference effect'' causes a slow down 
if $d_g<4d_w$.
\medskip

\begin{figure}
\includegraphics[width=\columnwidth]{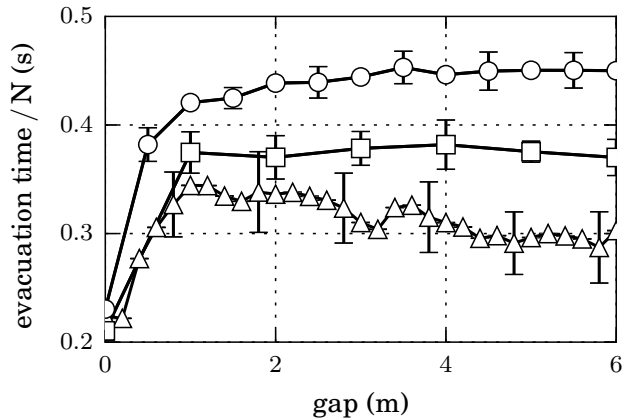}
\caption{\label{fig:1} Mean evacuation time per total number of pedestrians 
that left the room ($N$), as a function of the doors separation distance. Mean 
values were computed from 30 evacuation processes. Each door was $d_w=1.2$~m 
width for non-vanishing gaps. The null gap means a single door of $2L$ width. 
Three situations are shown: $\bigtriangleup$ corresponds to the $20\times20$~m 
room when 160 pedestrians left the room, $\Box$ corresponds to $30\times30$~m 
room when 530 pedestrians left the room, and $\bigcirc$ corresponds to 
$40\times40$~m room when  865 pedestrians left the room. The desired velocity 
was $v_d=4\,$m/s. }
% done with fig1_version0.py 
\end{figure}

The critical distance $4d_w$ has also been identified in the literature as the 
separation distance necessary to distinguish two independent pedestrian bulks 
around each door \cite{perez1}. We checked over this assertion by computing the 
mean evacuation time for an increasing number of pedestrians (and room sizes). 
We kept the pedestrian density unchanged (at $t=0$) for all the simulation 
processes. Fig.~\ref{fig:1} exhibits the mean evacuation time per pedestrian as 
a function of the separation distance (\emph{i.e.} gap). We divided the 
evacuation time by the total number of pedestrians for visualization 
reasons. 
\medskip

The results shown in Fig.~\ref{fig:1} were not expected. The evacuation time 
settles to an asymptotic value for separation distances $d_g>5\,$m. This means 
that the critical distance $4d_w$ and the bulk separation distance are actually 
not related, as proposed in Ref.~\cite{perez1}. The mean evacuation time 
becomes almost independent of the separation distances $d_g$ despite that the 
clogging areas around the doors might still overlap. 
\medskip 

Fig.~\ref{fig:1} also shows that the derivative not always changes sign  
at $d_g\simeq1\,$m. Furthermore, as the number of pedestrians is increased for 
$d_g>1\,$m, the evacuation time derivative raises to positive values. The 
greater the number of pedestrians, the worst evacuation time (per 
individual). This appears to occur for $d_g>1\,$m, regardless of the crowd 
size. That is, according to Fig.~\ref{fig:1}, there exists a separation 
distance value $d_g\simeq1\,$m where the evacuation derivative changes sharply 
to negative or positive values (for $d_g>1\,$m). This phenomenon has not been 
studied in the literature, to our knowledge. 
\medskip

We can resume the results in Fig.~\ref{fig:1} in the following way: the 
evacuation time rises when the doors separation increases from a wide opening 
(null separation distance) to the distance $d_g\simeq1\,$m. At this gap, the 
evacuation time derivative changes sharply, entering a much slowly 
varying regime towards an asymptotic value (for $d_g\gg1\,$m).  The former can 
be identified as a regime for small values of $d_g$, while the latter is valid 
for moderate to large values of $d_g$. The fact that a sharp change occurs at 
$d_g\simeq1\,$m, no matter the crowd size, suggests that both regimes are 
somehow different in nature. This moved us to explore the two regimes 
separately. 
\medskip

\subsubsection{\label{small_regime} The regime for $d_g<1\,$m}

\begin{figure*}[!htbp]
\subfloat[Separation distance of $1.5$~m.\label{fig:16}]{
\includegraphics[width=\columnwidth]{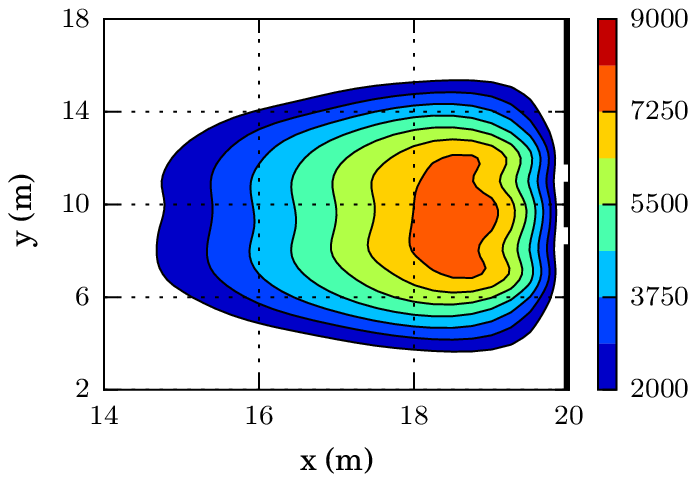}
% done with fig16_version0.py 
}\hfill
\subfloat[Separation distance of $5$~m. \label{fig:17}]{
\includegraphics[width=\columnwidth]{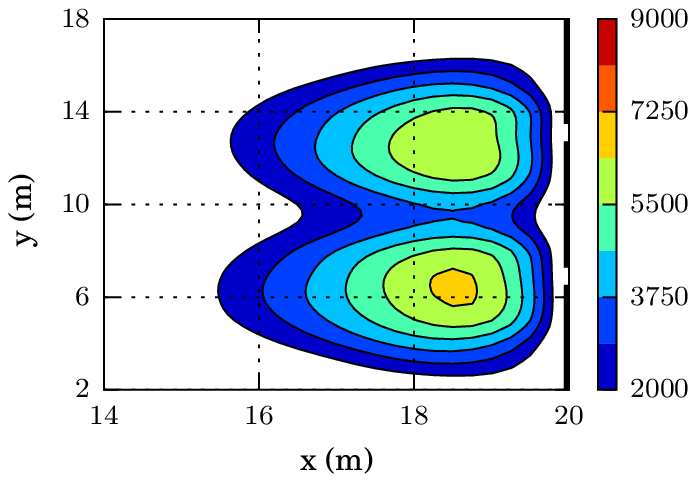}
% done with fig17_version0.py 
}
\caption{\label{fig:16and17} Mean pressure contour lines computed from 30 
evacuation processes until 100 pedestrians left the room 
($20\,\mathrm{m}\times20\,\mathrm{m}$ size). The scale bar on the right is 
expressed in N.m units (see text for details). The thick black lines at 
$x=20$~m represent the walls on the right of the room. The pedestrian's desired 
velocity was $v_d=4\,$m/s. The contour lines were computed on a square grid of 
$1\,\mathrm{m}\times1\,\mathrm{m}$ and then splined to get smooth curves. Level 
colors can be seen in the on-line version only.} 
\end{figure*}

Our starting point is the pressure contour map, since we can easily compare the 
current patterns with those presented in Section~\ref{null_gap_patterns}. 
Fig.~\ref{fig:16} shows the mean pressure pattern for the separation distance 
$d_g=1.5\,$m, that is, close to the gap value where the sharp change in the 
derivative occurs. The differences between Fig.~\ref{fig:16} and 
Fig.~{fig:2and4} are noticeable. We can now see a wide region in the center of 
the clogging area representing the high pressure ($3P_iV_i$) acting on each 
pedestrian (warm color in Fig.~\ref{fig:16}). The regularity in the color of 
this region is meaningful: the high pressure acting on the pedestrians does not 
allow a regular stream (pathway) to the exit. This is in agreement with the 
evacuation time worsening shown in Fig.~\ref{fig:13}.   
\medskip

\begin{figure}
\includegraphics[width=\columnwidth]{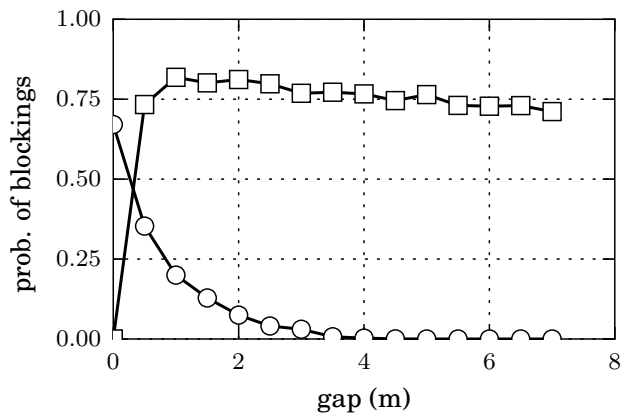}
\caption{\label{fig:14} Ratio between time steps including blocking structures 
and the total number of time steps for 30 evacuation processes, as a function 
of the doors separation distance. The room size was $20\times20$~m with 225 
occupants. Each door was $d_w=1.2$~m width for non-vanishing gaps. The null gap 
means a single door of $2d_w$ width. The desired velocity was $v_d=4\,$m/s.  
$\bigcirc$ corresponds blocking structures connecting both the left side wall 
of the left door with the right side wall of the right door (see text for 
details). $\Box$ corresponds to blocking structures connecting both sides of a 
single door (see text for details).   }
% done with fig14_version0.py 
\end{figure}

Fig.~\ref{fig:16} suggests that blocking structures might be present for long 
time periods, since the pedestrians cannot manage to get out easily. We 
examined this possibility through the \textit{blocking probability}. In this 
context, the blocking probability is associated to the ratio between the time 
that each door remains blocked with respect to the total evacuation time (cf. 
Section~\ref{human}). Fig.~\ref{fig:14} presents two kinds of blockings: the 
simultaneous blocking of both doors, and the blocking of a single door (say, 
the one on the left). The former connects the left most wall with the right 
most wall, but does not contact the separation wall in the middle of the walls. 
The latter connects the walls on both sides of the selected door (say, 
the one on the left). 
\medskip

According to Fig.~\ref{fig:14}, the single door blockings are not relevant 
until $d_g\simeq1\,$m, while the simultaneous blockings weaken as the gap 
(separation distance $d_g$) increases. The single door blockings resemble the 
response in Fig.~\ref{fig:13}, and thus, we conclude that this kind of 
blockings should play an important role in the increase of the evacuation time 
for small gaps $d_g$. Notice that single door blocking probability explains the 
75\% of the evacuation time, as can be seen in Fig.~\ref{fig:14}. 
\medskip

The results so far moved us to focus closer on the dynamics around each door. 
We watched many animations of the evacuation process for gap distances between 
the null separation to $d_g=1.5\,$m (not shown). We realized that single 
door blockings hold if the gap is large enough to stop at least two 
pedestrians. That is, any blocking structure enclosing a single door can hold 
for some time if the pedestrians at the end of the structure (and in contact 
with the walls) do hardy leave the structure. Two pedestrians are needed at the 
gap wall to ensure that both doors remain blocked.
\medskip

We want to call the attention on the fact that in the $d_g\simeq1\,$m scenario, 
the kind of simultaneous blocking without contacting the gap wall, 
is replaced by the kind of single door blockings acting (usually) 
simultaneously. This achieves a qualitative different pressure and 
stream pattern. As shown in Fig.~\ref{fig:2}, the widening of the exit 
allows a pathway through the middle of the clogging area. This is likely to 
occur even for very small gaps (see Fig.~\ref{fig:14}). However, the single door 
blockings follow a pressure pattern similar to  Fig.~\ref{fig:4} on each door.  
What we see in Fig.~\ref{fig:16} is the combined pattern built from two 
single door patterns as in Fig.~\ref{fig:4}. 
\medskip

We conclude from the analysis of small gaps ($d_g<1\,$m) that a door 
separation distance roughly equal to two pedestrian widths is critical. 
This distance allows persistent single door blockings. Small distances (close 
to the null separation) do not actually allow single door blockings to hold for 
long time. Thus, the role of $d_g=2d_w$ is decisive to move the evacuation 
process from one regime to another. 
\medskip

\subsubsection{\label{large_regime} The regime for $d_g>1\,$m}

\begin{figure}
\includegraphics[width=\columnwidth]{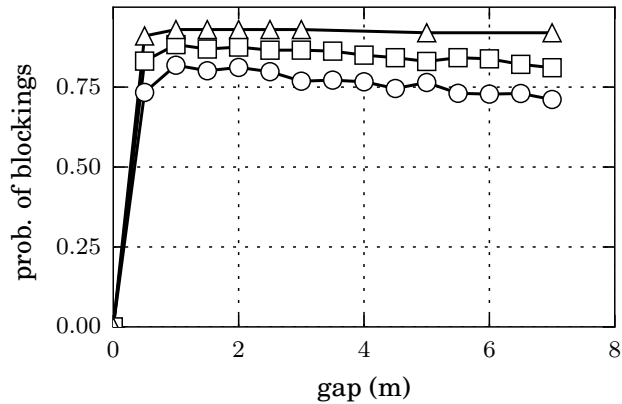}
\caption{\label{fig:15} Ratio between time steps including blocking structures 
and the total number of time steps for 30 evacuation processes, as a function 
of the doors separation distance. The only blocking structures considered 
were those connecting both sides of one single door (see text for details). 
Each door was $d_w=1.2$~m width for non-vanishing gaps. The null gap means a 
single door of $2d_w$ width. Three scenarios are shown: $\bigcirc$ 
corresponds to the room of size $20\times20$~m with 225 occupants and a 
desired velocity of $v_d=4\,$m/s. $\Box$ corresponds to the room of size  
$20\times20$~m with 225 occupants and a desired velocity of $v_d=6\,$m/s. 
$\bigtriangleup$ corresponds to the room of size $40\times40$~m with 961 
occupants and a desired velocity of $v_d=4\,$m/s.   }
% done with fig15_version0.py 
\end{figure}

Fig.~\ref{fig:14} shows that the single door blockings (see 
 Section~\ref{small_regime}) remains around 75\% of the total evacuation time 
for $d_g>1\,$m (225 individuals in the room). We also computed this magnitude 
for situations with increasing number of individuals (see Fig.~\ref{fig:15}). 
The probability of single door blockings approaches unity as the crowd size 
increases. This means, according to our definition of blocking probability, 
that the blocking time raises as the number of individuals increases. The gap 
distance, however, does not play a role for $d_g>1\,$m.  
\medskip

There is a noticeable difference between the evacuation time 
shown in Fig.~\ref{fig:1} and the blocking probability exhibited in 
Fig.~\ref{fig:15}. The derivative changes sign in the former (for increasing 
number of pedestrians), but it does not in the latter. Therefore, the 
blocking time cannot be considered as the reason for this changes.   \medskip

We examined many animations of the evacuation process for an increasing number 
of pedestrians and separation distances. We also checked the pressure patterns 
for $d_g>4d_w$ (see Fig.~\ref{fig:17} as an example). We came to the conclusion 
that since the evacuation derivative in Fig.~\ref{fig:1} changes with an
increasing number of individuals, the whole bulk should be involved in 
this phenomenon. Therefore, we focused our investigation on the pressure 
contribution of the whole bulk.   
\medskip

Eq.~\ref{eqn_5} relates the ``social pressure function'' (left-hand side) with 
the desire force contribution (right-hand side). That is, an increase in the 
desire force of the individuals (\emph{i.e.} anxiety levels) means an increase 
in the bulk ``social pressure''. A simple example on the Eq.~\ref{eqn_5} 
computation can be found in the Appendix bellow.   
\medskip

Fig.~\ref{fig:12} shows the evacuation time as a function of the separation 
distance for two different desired velocities. As expected, the sharp change in 
the derivative occur around $d_g=2d_w$. Also the derivative changes as the 
desired velocity is increased (\emph{i.e.} higher anxiety level). This confirms 
that the social pressure is responsible the derivative behaviour shown in  
Fig.~\ref{fig:1}. 
\medskip

\begin{figure}
\includegraphics[width=\columnwidth]{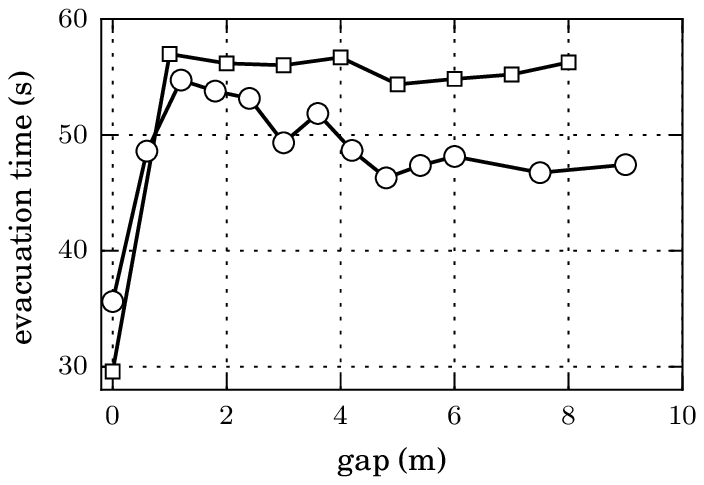}
\caption{\label{fig:12} Mean evacuation time for 225 pedestrians (room of 
$20\times20$~m size) as a function of the doors separation distance. Mean 
values were computed from 30 evacuation processes until 160 pedestrians left 
the room. Each door was $d_w=1.2$~m width for non-vanishing gaps. The null gap 
means a single door of $2L$ width. $\bigcirc$ corresponds to pedestrians 
with desired velocity of $v_d=4\,$m/s. $\Box$ corresponds to pedestrians 
with desired velocity of $v_d=8\,$m/s. }
% done with fig12_version0.py 
\end{figure}

We conclude from the analysis of large gaps ($d_g>1\,$m) that the evacuation 
time is controlled by the social pressure in the bulk. The crowd size and 
the desired velocity $v_d$ affects the pressure acting on the pedestrians. But 
no further changes in the evacuation time can be noticed for$d_g>4d_w$. This 
means that the asymptotic evacuation value does not depend strongly on 
weather the bulks around each door are completely independent.

\section{\label{conclusions}Conclusions}

We examined in detail the evacuation of pedestrians for the situation where 
two contiguous doors are available for leaving the room. Throughout  
Section~\ref{results} we presented results on the evacuation performance under 
high anxiety levels and increasing number of pedestrians. Both conditions 
exhibit the novel result that a worsening in the evacuation time as the door 
separation distance $d_g$ increases from the null value to roughly the width of 
two pedestrians. Special situations may enhance the evacuation performance for 
larger values of $d_g$. 
\medskip

Two regimes were identified as the $d_g$ values increased from $d_g=0$ to 
$d_g>d_w$ (2 pedestrians width). The range $0\leq d_g\leq 2d_w$ worsened the 
evacuation performance for all the explored situations, while the 
range $d_g>2d_w$ did enhance the evacuation time for relatively small crowds 
and moderate anxiety levels. We realized that the sharp change in the 
evacuation behaviour at $d_g=2d_w$ corresponded to qualitative differences in 
the pedestrian dynamics close to the exits.
\medskip

After a detailed comparison of the dynamics for the single door situation and 
for two doors very close to each other (that is $d_g<2d_w$), we concluded that 
the blocking structures (\emph{i.e.} blocking archs) around the openings were 
released intermittently, allowing the pedestrians to leave the room in a 
stop-and-go process. But, as the separation distance approached $2d_w$,  the 
blocking archs were restored around each door, resembling the blocking 
situation of two a single doors. This changes only affected the 
local dynamics (close to the doors), while the crowd remained gathered into a 
single clogging area. 
\medskip

Starting at $d_g=2d_w$ allows the single door blocking structures to become 
relevant even for large values of $d_g$ (see Fig.\ref{fig:14}). No further 
qualitative changes were 
observed locally around each door. However, increasing the crowd size ($N$) or 
the pedestrian's anxiety level ($v_d$) slowed down the evacuation. Both 
magnitudes are linked to the pressure acting on the pedestrians, and therefore, 
enhanced the ``faster is slower'' affects. 
\medskip

For a better understanding of the relationship between $N$, $v_d$ and the 
pressure in the bulk, a simple lane example complemented our analysis. It was 
shown that the classical virial expression is still suitable for the 
investigation of social systems.

\begin{acknowledgments}
C.O.~Dorso is a main researcher of the National Scientific and Technical 
Research Council (spanish: Consejo Nacional de Investigaciones Cient\'\i ficas y 
T\'ecnicas - CONICET), Argentina. G.A.~Frank is an assistant researcher of the 
CONICET, Argentina. I.M.~Sticco has degree in Physics.
\end{acknowledgments}

\appendix

\section{\label{app}The lane example}

We decided to open this supplementary section in order to make clear the 
meaning of the ``social pressure'' acting on an individual and the collective 
pressure (that is, the \textit{bulk} pressure) on a set of individuals. We 
will follow a simple example as a guide for more general situations. \medskip

\subsection{\label{social_pressure}The social pressure}

\begin{figure}[!htbp]
\includegraphics[width=0.75\columnwidth]{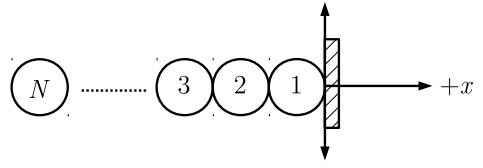}
\caption{\label{fig:18} Lane of individuals pushing to the right. The 
horizontal axis indicates the positive direction.  }
% done with figuras_presion.odg
\end{figure}

Fig.~\ref{fig:18} represents a lane of individuals pushing to the right. The 
ending wall prevents the individuals from moving. All the pedestrians in the 
lane are at their equilibrium positions $x_1,x_2,...,x_{i},...x_N$, 
while the wall is placed at the position $x_0=0$ (see Fig.~\ref{fig:18}).  
\medskip

The pedestrians push to the right acknowledging a desired force 
$f_d^{(i)}=mv_d/\tau$, according to Eq.~(\ref{eqn_2}). The social repulsion 
feelings balance this desire force, but only the contacting neighbors are 
relevant to these feelings. Thus, the balance equation for any pedestrian 
in the lane reads 
\medskip

\begin{equation}
 f_s^{(i,i+1)}-f_s^{(i,i-1)}+\displaystyle\frac{mv_d}{\tau}=0\label{eqn_6}
\end{equation}

\noindent for $f_s^{(i,j)}$ meaning the repulsive feelings of pedestrian $i$ 
due to the presence of pedestrian $j$. Notice that the boundary condition at 
the wall-end is $x_0=0$ (Dirichlet condition), while the condition at the free 
end is $f_s^{(N,N+1)}=0$ (Neumann condition). The forces on the pedestrians can 
be obtained recursively from Eq.~\ref{eqn_6}, starting at the free ended 
individual ($i=N$). The resulting expression is 

\begin{equation}
f_s^{(i,i-1)}=(N-i+1)\,\displaystyle\frac{mv_d}{\tau}\ \ \ , \ \ \ 
i=1,....,N\label{eqn_7}
\end{equation}

\noindent while the corresponding positions $x_1,x_2,...,x_{i},...x_N$ are 
obtained by a backward substitution of the social forces expressed in 
Eq.~\ref{eqn_2}, starting at the wall-end

\begin{equation} 
x_i=x_{i-1}-(r_{i}+r_{i-1})+B\,\ln\bigg[(N-i+1)\,\displaystyle\frac{mv_d}{A\tau}
\bigg]\label{eqn_8}
\end{equation}

Our intuition suggests that the pressure on a single pedestrian $P_i$ 
corresponds to the forces acting on him (her) (per unit area) due to the 
neighboring pedestrians. We can further assert this from the 
``social pressure function'' definition (\ref{eqn_4})  

\begin{equation}
P_i=\displaystyle\frac{1}{2}\,\bigg[\displaystyle\frac{x_{i}-x_{i+1}}{3V_i}\,
f_s^ { (i , i+1) } +\displaystyle\frac { x_ {i-1}-x_{i}}{3V_i}\,f_s^{(i,i-1) 
}\bigg]\label{eqn_9}
\end{equation}

\vspace{3mm}

\noindent where the magnitude $x_{ij}/3V_i$ corresponds to the 
(inverse) effective surface of the pedestrian. For individuals modeled as 
hard spheres, the inter-pedestrian distance is $x_{ij}=2r_i$ and the volume 
is $V_i=4\pi r_i^3/3$. Thus, 

\begin{equation}
P_i=\displaystyle\frac{1}{4\pi 
r_i^2}\,\bigg[f_s^ { (i , i+1) } +f_s^{(i,i-1)}\bigg]\label{eqn_10}
\end{equation}

\noindent as expected for the individual pressure.  
\medskip

\subsection{\label{bulk_pressure}The bulk pressure}

We can first check over the virial relation (\ref{eqn_5}) through the 
expression (\ref{eqn_9}). Adding the terms for the lane of $N$ pedestrians and 
replacing the first and last term with the corresponding boundary condition, 
gives

\begin{equation}
\left\{\begin{array}{lcl}
3P_1V_1 
& = &\displaystyle\frac{x_{1}}{2}f_s^ { (1 , 2)} - 
\displaystyle\frac{x_{2}}{2}\,f_s^ { (1 , 2) }  \\
&& \\
3P_2V_2 
& = &\displaystyle\frac{x_{2}}{2}\, \big[f_s^ { (2 , 3)} - f_s^{(2,1)} 
\big] - \\
&& \displaystyle\frac{x_{3}}{2}\,f_s^ { (2 , 3) } +\displaystyle\frac 
{ x_ {1}}{2}\,f_s^{(2,1) 
} \\
&& \\
3P_3V_3 
& = &\displaystyle\frac{x_{3}}{2}\,\big[f_s^ { (3 , 4)} - f_s^{(3,2)} 
\big]- \\
&& \displaystyle\frac{x_{4}}{2}\,f_s^ { (3 , 4) } +\displaystyle\frac 
{ x_{2}}{2}\,f_s^{(3,2) 
} \\
... &&\\
&& \\
3P_NV_N 
& = &-\displaystyle\frac{x_{N}}{2}\, f_s^{(N,N-1)} 
+ \\
&& \displaystyle\frac{ x_{N-1}}{2}\,f_s^{(N,N-1) 
} \\
 \end{array}\right.\label{eqn_11}
\end{equation}
\medskip

These are the local pressures on each pedestrian due to the 
contacting neighbors (and excluding the wall). Adding the terms results 
in the virial relation, as expressed in (\ref{eqn_5})

\begin{equation}
\begin{array}{lcl}
\displaystyle\sum_{i=1}^N 3P_iV_i & = & (x_1 - x_2)f_s^{(1,2)} + (x_2 - 
x_3)f_s^{(2,3)} +... \\
& + &  (x_{N-1}-x_N)f_s^{(N,N-1)} \\
&& \\
& = & x_{1}\,\displaystyle\frac{N\,mv_d}{\tau} - 
\displaystyle\sum_{i=1}^N x_i\,\displaystyle\frac{mv_d}{\tau} \\
 \end{array}\label{eqn_12}
\end{equation}

\noindent where the first term on the right corresponds to the global pressure 
$-3\mathcal{PV}$. Notice that $x_1$ is negative, and thus, $3\mathcal{PV}$ is 
defined as a positive magnitude. The last term is also positive, adding 
pressure to the bulk due to the desire forces.  
\medskip

The virial relation (\ref{eqn_5}) allows to compute the \textit{bulk} pressure 
on a group of pedestrians. For example, the pressure on the $M$ pedestrians 
closest to the wall corresponds to the force acting on this group due to the 
other $N-M$ pedestrians. According to Eq.~(\ref{eqn_5}), the 
pressure on the $M$ individuals is

\begin{equation}
 \displaystyle\sum_{i=1}^M 3P_iV_i 
=-3\mathcal{PV}-\displaystyle\sum_{i=M+1}^N 3P_iV_i-\displaystyle\sum_{i=1}^N 
x_i\displaystyle\frac{mv_d}{\tau}\label{eqn_13}
\end{equation}

The \textit{bulk} pressure on the first $M$ individuals increases as more 
individuals are included in the crowd. This can be verified by evaluating 
Eq.~(\ref{eqn_12}) and Eq.~(\ref{eqn_13}) for increasing values of $N$. 
\medskip

\begin{figure}[!htbp]
\includegraphics[width=\columnwidth]{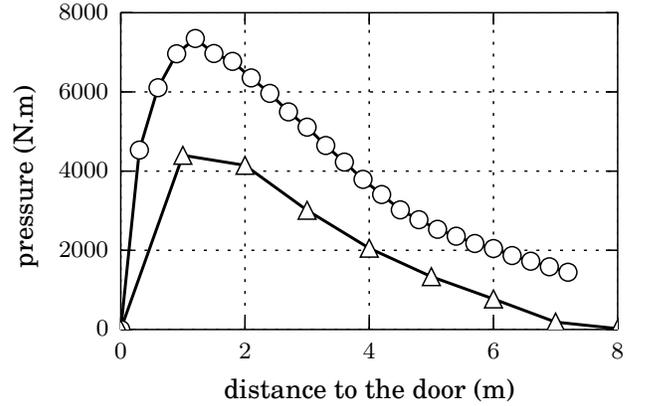}
\caption{\label{fig:10} Mean pressure as a function of the distance 
to the exit. The room was $20\,\mathrm{m}\times20\,\mathrm{m}$ size and 
included one door of $d_w=1.2$~m width. Mean values were computed from 30 
evacuation processes, until 100 pedestrians left the room. The desired velocity 
was $v_d=4\,$m/s. The distance to the door was binned into equal intervals of 
$0.3$~m or 1~m.  The $\bigcirc$  symbols correspond to bins of $0.3$~m long. 
The 
symbols $\bigtriangleup$ correspond to bins of 1~m long, but the recording was 
restricted to $9.5\,\mathrm{m}\leq y\leq 10.5\,\mathrm{m}$ (see text for 
details).  }
% done with fig10_version0.py 
\end{figure}

The Eqs.~(\ref{eqn_7}) and (\ref{eqn_8}) allow to compute the pedestrian 
pressure profile as a function of the distance to the wall. The profile is 
qualitatively similar to the one measured during an evacuation process. 
Fig.~\ref{fig:10} represents the histogram for the pressure on each 
pedestrian.

\bibliography{paper}% Produces the bibliography via BibTeX.

%merlin.mbs apsrev4-1.bst 2010-07-25 4.21a (PWD, AO, DPC) hacked
%Control: key (0)
%Control: author (8) initials jnrlst
%Control: editor formatted (1) identically to author
%Control: production of article title (-1) disabled
%Control: page (0) single
%Control: year (1) truncated
%Control: production of eprint (0) enabled
\begin{thebibliography}{17}%
\makeatletter
\providecommand \@ifxundefined [1]{%
 \@ifx{#1\undefined}
}%
\providecommand \@ifnum [1]{%
 \ifnum #1\expandafter \@firstoftwo
 \else \expandafter \@secondoftwo
 \fi
}%
\providecommand \@ifx [1]{%
 \ifx #1\expandafter \@firstoftwo
 \else \expandafter \@secondoftwo
 \fi
}%
\providecommand \natexlab [1]{#1}%
\providecommand \enquote  [1]{``#1''}%
\providecommand \bibnamefont  [1]{#1}%
\providecommand \bibfnamefont [1]{#1}%
\providecommand \citenamefont [1]{#1}%
\providecommand \href@noop [0]{\@secondoftwo}%
\providecommand \href [0]{\begingroup \@sanitize@url \@href}%
\providecommand \@href[1]{\@@startlink{#1}\@@href}%
\providecommand \@@href[1]{\endgroup#1\@@endlink}%
\providecommand \@sanitize@url [0]{\catcode `\\12\catcode `\$12\catcode
  `\&12\catcode `\#12\catcode `\^12\catcode `\_12\catcode `\%12\relax}%
\providecommand \@@startlink[1]{}%
\providecommand \@@endlink[0]{}%
\providecommand \url  [0]{\begingroup\@sanitize@url \@url }%
\providecommand \@url [1]{\endgroup\@href {#1}{\urlprefix }}%
\providecommand \urlprefix  [0]{URL }%
\providecommand \Eprint [0]{\href }%
\providecommand \doibase [0]{http://dx.doi.org/}%
\providecommand \selectlanguage [0]{\@gobble}%
\providecommand \bibinfo  [0]{\@secondoftwo}%
\providecommand \bibfield  [0]{\@secondoftwo}%
\providecommand \translation [1]{[#1]}%
\providecommand \BibitemOpen [0]{}%
\providecommand \bibitemStop [0]{}%
\providecommand \bibitemNoStop [0]{.\EOS\space}%
\providecommand \EOS [0]{\spacefactor3000\relax}%
\providecommand \BibitemShut  [1]{\csname bibitem#1\endcsname}%
\let\auto@bib@innerbib\@empty
%</preamble>
\bibitem [{\citenamefont {Cheng}\ \emph {et~al.}(2004)\citenamefont {Cheng},
  \citenamefont {Lo}, \citenamefont {Fang},\ and\ \citenamefont
  {Cheng}}]{cheng}%
  \BibitemOpen
  \bibfield  {author} {\bibinfo {author} {\bibfnamefont {W.}~\bibnamefont
  {Cheng}}, \bibinfo {author} {\bibfnamefont {S.}~\bibnamefont {Lo}}, \bibinfo
  {author} {\bibfnamefont {Z.}~\bibnamefont {Fang}}, \ and\ \bibinfo {author}
  {\bibfnamefont {C.}~\bibnamefont {Cheng}},\ }\href@noop {} {\bibfield
  {journal} {\bibinfo  {journal} {Structural Survey}\ }\textbf {\bibinfo
  {volume} {22}},\ \bibinfo {pages} {201–209} (\bibinfo {year}
  {2004})}\BibitemShut {NoStop}%
\bibitem [{\citenamefont {OSHA}(2015)}]{OSHA}%
  \BibitemOpen
  \bibfield  {author} {\bibinfo {author} {\bibnamefont {OSHA}},\ }\href@noop {}
  {\bibfield  {journal} {\bibinfo  {journal} {Occupational Safety \& Health
  Administration Standards 29-CFR}\ }\textbf {\bibinfo {volume} {1910.36(b)}},\
  \bibinfo {pages} {1} (\bibinfo {year} {2015})}\BibitemShut {NoStop}%
\bibitem [{\citenamefont {FBC2010}(2010{\natexlab{a}})}]{FLO}%
  \BibitemOpen
  \bibfield  {author} {\bibinfo {author} {\bibnamefont {FBC2010}},\ }\href@noop
  {} {\bibfield  {journal} {\bibinfo  {journal} {Florida Building Code
  Handbook}\ }\textbf {\bibinfo {volume} {1015.1}},\ \bibinfo {pages} {90}
  (\bibinfo {year} {2010}{\natexlab{a}})}\BibitemShut {NoStop}%
\bibitem [{\citenamefont {FBC2010}(2010{\natexlab{b}})}]{FLO2}%
  \BibitemOpen
  \bibfield  {author} {\bibinfo {author} {\bibnamefont {FBC2010}},\ }\href@noop
  {} {\bibfield  {journal} {\bibinfo  {journal} {Florida Building Code
  Handbook}\ }\textbf {\bibinfo {volume} {1008.1}},\ \bibinfo {pages} {81}
  (\bibinfo {year} {2010}{\natexlab{b}})}\BibitemShut {NoStop}%
\bibitem [{\citenamefont {Kirchner}\ and\ \citenamefont
  {Schadschneider}(2002)}]{kirchner1}%
  \BibitemOpen
  \bibfield  {author} {\bibinfo {author} {\bibfnamefont {A.}~\bibnamefont
  {Kirchner}}\ and\ \bibinfo {author} {\bibfnamefont {A.}~\bibnamefont
  {Schadschneider}},\ }\href@noop {} {\bibfield  {journal} {\bibinfo  {journal}
  {Physica A}\ }\textbf {\bibinfo {volume} {312}},\ \bibinfo {pages} {260}
  (\bibinfo {year} {2002})}\BibitemShut {NoStop}%
\bibitem [{\citenamefont {Perez}\ \emph {et~al.}(2002)\citenamefont {Perez},
  \citenamefont {Tapang}, \citenamefont {Lim},\ and\ \citenamefont
  {Saloma}}]{perez1}%
  \BibitemOpen
  \bibfield  {author} {\bibinfo {author} {\bibfnamefont {G.}~\bibnamefont
  {Perez}}, \bibinfo {author} {\bibfnamefont {G.}~\bibnamefont {Tapang}},
  \bibinfo {author} {\bibfnamefont {M.}~\bibnamefont {Lim}}, \ and\ \bibinfo
  {author} {\bibfnamefont {C.}~\bibnamefont {Saloma}},\ }\href@noop {}
  {\bibfield  {journal} {\bibinfo  {journal} {Physica A}\ }\textbf {\bibinfo
  {volume} {312}},\ \bibinfo {pages} {609} (\bibinfo {year}
  {2002})}\BibitemShut {NoStop}%
\bibitem [{\citenamefont {Daoliang}\ \emph {et~al.}(2006)\citenamefont
  {Daoliang}, \citenamefont {Lizhong},\ and\ \citenamefont {Jian}}]{daoliang1}%
  \BibitemOpen
  \bibfield  {author} {\bibinfo {author} {\bibfnamefont {Z.}~\bibnamefont
  {Daoliang}}, \bibinfo {author} {\bibfnamefont {Y.}~\bibnamefont {Lizhong}}, \
  and\ \bibinfo {author} {\bibfnamefont {L.}~\bibnamefont {Jian}},\ }\href@noop
  {} {\bibfield  {journal} {\bibinfo  {journal} {Physica A}\ }\textbf {\bibinfo
  {volume} {363}},\ \bibinfo {pages} {501} (\bibinfo {year}
  {2006})}\BibitemShut {NoStop}%
\bibitem [{\citenamefont {Huan-Huan}\ \emph {et~al.}(2015)\citenamefont
  {Huan-Huan}, \citenamefont {Li-Yun},\ and\ \citenamefont {Yu}}]{huanhuan1}%
  \BibitemOpen
  \bibfield  {author} {\bibinfo {author} {\bibfnamefont {T.}~\bibnamefont
  {Huan-Huan}}, \bibinfo {author} {\bibfnamefont {D.}~\bibnamefont {Li-Yun}}, \
  and\ \bibinfo {author} {\bibfnamefont {X.}~\bibnamefont {Yu}},\ }\href@noop
  {} {\bibfield  {journal} {\bibinfo  {journal} {Physica A}\ }\textbf {\bibinfo
  {volume} {420}},\ \bibinfo {pages} {164} (\bibinfo {year}
  {2015})}\BibitemShut {NoStop}%
\bibitem [{\citenamefont {Helbing}\ \emph {et~al.}(2000)\citenamefont
  {Helbing}, \citenamefont {Farkas},\ and\ \citenamefont {Vicsek}}]{Helbing1}%
  \BibitemOpen
  \bibfield  {author} {\bibinfo {author} {\bibfnamefont {D.}~\bibnamefont
  {Helbing}}, \bibinfo {author} {\bibfnamefont {I.}~\bibnamefont {Farkas}}, \
  and\ \bibinfo {author} {\bibfnamefont {T.}~\bibnamefont {Vicsek}},\
  }\href@noop {} {\bibfield  {journal} {\bibinfo  {journal} {Nature}\ }\textbf
  {\bibinfo {volume} {407}},\ \bibinfo {pages} {487} (\bibinfo {year}
  {2000})}\BibitemShut {NoStop}%
\bibitem [{\citenamefont {Parisi}\ and\ \citenamefont {Dorso}(2005)}]{Dorso1}%
  \BibitemOpen
  \bibfield  {author} {\bibinfo {author} {\bibfnamefont {D.}~\bibnamefont
  {Parisi}}\ and\ \bibinfo {author} {\bibfnamefont {C.}~\bibnamefont {Dorso}},\
  }\href@noop {} {\bibfield  {journal} {\bibinfo  {journal} {Physica A}\
  }\textbf {\bibinfo {volume} {354}},\ \bibinfo {pages} {606} (\bibinfo {year}
  {2005})}\BibitemShut {NoStop}%
\bibitem [{\citenamefont {Parisi}\ and\ \citenamefont {Dorso}(2007)}]{Dorso2}%
  \BibitemOpen
  \bibfield  {author} {\bibinfo {author} {\bibfnamefont {D.}~\bibnamefont
  {Parisi}}\ and\ \bibinfo {author} {\bibfnamefont {C.}~\bibnamefont {Dorso}},\
  }\href@noop {} {\bibfield  {journal} {\bibinfo  {journal} {Physica A}\
  }\textbf {\bibinfo {volume} {385}},\ \bibinfo {pages} {343} (\bibinfo {year}
  {2007})}\BibitemShut {NoStop}%
\bibitem [{\citenamefont {Frank}\ and\ \citenamefont {Dorso}(2011)}]{Dorso3}%
  \BibitemOpen
  \bibfield  {author} {\bibinfo {author} {\bibfnamefont {G.}~\bibnamefont
  {Frank}}\ and\ \bibinfo {author} {\bibfnamefont {C.}~\bibnamefont {Dorso}},\
  }\href@noop {} {\bibfield  {journal} {\bibinfo  {journal} {Physica A}\
  }\textbf {\bibinfo {volume} {390}},\ \bibinfo {pages} {2135} (\bibinfo {year}
  {2011})}\BibitemShut {NoStop}%
\bibitem [{\citenamefont {Frank}\ and\ \citenamefont {Dorso}(2015)}]{Dorso4}%
  \BibitemOpen
  \bibfield  {author} {\bibinfo {author} {\bibfnamefont {G.}~\bibnamefont
  {Frank}}\ and\ \bibinfo {author} {\bibfnamefont {C.}~\bibnamefont {Dorso}},\
  }\href@noop {} {\bibfield  {journal} {\bibinfo  {journal} {International
  Journal of Modern Physics C}\ }\textbf {\bibinfo {volume} {26}},\ \bibinfo
  {pages} {1} (\bibinfo {year} {2015})}\BibitemShut {NoStop}%
\bibitem [{\citenamefont {Helbing}\ and\ \citenamefont
  {Moln\'ar}(1995)}]{Helbing4}%
  \BibitemOpen
  \bibfield  {author} {\bibinfo {author} {\bibfnamefont {D.}~\bibnamefont
  {Helbing}}\ and\ \bibinfo {author} {\bibfnamefont {P.}~\bibnamefont
  {Moln\'ar}},\ }\href@noop {} {\bibfield  {journal} {\bibinfo  {journal}
  {Physical Review E}\ }\textbf {\bibinfo {volume} {51}},\ \bibinfo {pages}
  {4282} (\bibinfo {year} {1995})}\BibitemShut {NoStop}%
\bibitem [{\citenamefont {Lion}\ and\ \citenamefont {Allen}(2012)}]{lion}%
  \BibitemOpen
  \bibfield  {author} {\bibinfo {author} {\bibfnamefont {T.~W.}\ \bibnamefont
  {Lion}}\ and\ \bibinfo {author} {\bibfnamefont {R.~J.}\ \bibnamefont
  {Allen}},\ }\href {http://stacks.iop.org/0953-8984/24/i=28/a=284133}
  {\bibfield  {journal} {\bibinfo  {journal} {Journal of Physics: Condensed
  Matter}\ }\textbf {\bibinfo {volume} {24}},\ \bibinfo {pages} {284133}
  (\bibinfo {year} {2012})}\BibitemShut {NoStop}%
\bibitem [{\citenamefont {Mysen}\ \emph {et~al.}(2005)\citenamefont {Mysen},
  \citenamefont {Berntsen}, \citenamefont {Nafstad},\ and\ \citenamefont
  {Schild}}]{mysen}%
  \BibitemOpen
  \bibfield  {author} {\bibinfo {author} {\bibfnamefont {M.}~\bibnamefont
  {Mysen}}, \bibinfo {author} {\bibfnamefont {S.}~\bibnamefont {Berntsen}},
  \bibinfo {author} {\bibfnamefont {P.}~\bibnamefont {Nafstad}}, \ and\
  \bibinfo {author} {\bibfnamefont {P.~G.}\ \bibnamefont {Schild}},\ }\href
  {\doibase http://dx.doi.org/10.1016/j.enbuild.2005.01.003} {\bibfield
  {journal} {\bibinfo  {journal} {Energy and Buildings}\ }\textbf {\bibinfo
  {volume} {37}},\ \bibinfo {pages} {1234 } (\bibinfo {year}
  {2005})}\BibitemShut {NoStop}%
\bibitem [{\citenamefont {Plimpton}(1995)}]{plimpton}%
  \BibitemOpen
  \bibfield  {author} {\bibinfo {author} {\bibfnamefont {S.}~\bibnamefont
  {Plimpton}},\ }\href {\doibase http://dx.doi.org/10.1006/jcph.1995.1039}
  {\bibfield  {journal} {\bibinfo  {journal} {Journal of Computational
  Physics}\ }\textbf {\bibinfo {volume} {117}},\ \bibinfo {pages} {1 }
  (\bibinfo {year} {1995})}\BibitemShut {NoStop}%
\end{thebibliography}%

\end{document}